\documentclass[apsrev4-1,twocolumn,footinbib,superscriptaddress,floatfix,bibliography]{revtex4-1}

\usepackage{times}
\usepackage{float}
\usepackage{multirow}
\usepackage[dvipsnames]{xcolor}
\usepackage{amsmath}
\usepackage{amsthm}
\usepackage{amssymb}
\usepackage{amsbsy}
\usepackage{enumitem}
\usepackage{wasysym}
\usepackage[english]{babel}
\usepackage[T1]{fontenc}
\usepackage[utf8]{inputenc} 
\usepackage{graphicx}
\usepackage[colorlinks,bookmarks=false,citecolor=blue,linkcolor=red,urlcolor=blue]{hyperref}
\usepackage{ulem}
\usepackage{pstricks}
\usepackage{rotating}			       
\usepackage{tabularx,hhline}	
\usepackage[caption=false]{subfig}	
\newcolumntype{P}[1]{>{\centering\arraybackslash}p{#1}}

\usepackage{tikz}
\usepackage{etoolbox}
\newrobustcmd*{\mysquare}[1]{\tikz{\filldraw[draw=#1,fill=#1] (0,0) rectangle (0.2cm,0.2cm);}}
\newrobustcmd*{\mycircle}[1]{\tikz{\filldraw[draw=#1,fill=#1] (0,0) circle [radius=0.1cm];}}
\newrobustcmd*{\mytriangle}[1]{\tikz{\filldraw[draw=#1,fill=#1] (0,0) -- (0.2cm,0) -- (0.1cm,0.2cm);}}
\newrobustcmd*{\myinvtriangle}[1]{\tikz{\filldraw[draw=#1,fill=#1] (0cm,0.2cm) -- (0.2cm,0.2cm) -- (0.1cm,0.0cm);}}
\newrobustcmd*{\mydiamond}[1]{\tikz{\filldraw[draw=#1,fill=#1] (0,0) -- (0.1cm,0.1cm) -- (0.0cm,0.2cm) -- (-0.1cm, 0.1cm); }}
\newrobustcmd*{\mypentagon}[1]{\tikz{\filldraw[draw=#1,fill=#1] (0,0) -- (0.05cm,0.0cm)--(0.1cm,0.1cm) -- (0.0cm,0.2cm) -- (-0.1cm, 0.1cm)--(-0.05cm,0.0cm)--(0.0cm,0.0cm);}}
\newrobustcmd*{\mysquaree}[1]{\tikz{\filldraw[draw=#1,fill=white, line width=0.7mm ] (0,0) rectangle (0.2cm,0.2cm) }}
\newrobustcmd*{\mycirclee}[1]{\tikz{\filldraw[draw=#1,fill=white, line width=0.7mm] (0,0) circle [radius=0.1cm];}}
\newrobustcmd*{\mytrianglee}[1]{\tikz{\filldraw[draw=#1,fill=white, line width=0.7mm] (0,0) -- (0.2cm,0) -- (0.1cm,0.2cm)--(0,0)--(0.2cm,0);}}
\newrobustcmd*{\myinvtrianglee}[1]{\tikz{\filldraw[draw=#1,fill=white, line width=0.7mm] (0cm,0.2cm) -- (0.2cm,0.2cm) -- (0.1cm,0.0cm)--(0cm,0.2cm);}}
\newrobustcmd*{\mydiamondd}[1]{\tikz{\filldraw[draw=#1,fill=white, line width=0.7mm] (0,0) -- (0.1cm,0.1cm) -- (0.0cm,0.2cm) -- (-0.1cm, 0.1cm)--(0,0)--(0.1cm,0.1cm); }}
\newrobustcmd*{\mypentagonn}[1]{\tikz{\filldraw[draw=#1,fill=white, line width=0.7mm] (0,0) -- (0.05cm,0.0cm)--(0.1cm,0.1cm) -- (0.0cm,0.2cm) -- (-0.1cm, 0.1cm)--(-0.05cm,0.0cm)--(0.0cm,0.0cm);}}
\newrobustcmd*{\mysquareee}[1]{\tikz{\filldraw[draw=#1,fill=white] (0,0) rectangle (0.2cm,0.2cm);}}
\newrobustcmd*{\mycircleee}[1]{\tikz{\filldraw[draw=#1,fill=white] (0,0) circle [radius=0.1cm];}}
\newrobustcmd*{\mytriangleee}[1]{\tikz{\filldraw[draw=#1,fill=white] (0,0) -- (0.2cm,0) -- (0.1cm,0.2cm)--(0cm,0cm);}}
\newrobustcmd*{\myinvtriangleee}[1]{\tikz{\filldraw[draw=#1,fill=white] (0cm,0.2cm) -- (0.2cm,0.2cm) -- (0.1cm,0.0cm) --(0cm,0.2cm);}}
\newrobustcmd*{\mydiamonddd}[1]{\tikz{\filldraw[draw=#1,fill=white] (0,0) -- (0.1cm,0.1cm) -- (0.0cm,0.2cm) -- (-0.1cm, 0.1cm)--(0,0); }}
\newrobustcmd*{\mypentagonnn}[1]{\tikz{\filldraw[draw=#1,fill=white] (0,0) -- (0.05cm,0.0cm)--(0.1cm,0.1cm) -- (0.0cm,0.2cm) -- (-0.1cm, 0.1cm)--(-0.05cm,0.0cm)--(0.0cm,0.0cm);}}
\definecolor{lightmagenta}{rgb}{0.9375,0.33203125,0.9375}
\definecolor{lightcoral}{rgb}{0.9375,0.5,0.5}
\definecolor{forestgreen(web)}{rgb}{0.13, 0.55, 0.13}
\definecolor{lightgreen}{rgb}{0.56, 0.93, 0.56}
\definecolor{royalblue(web)}{rgb}{0.25, 0.41, 0.88}
\definecolor{skyblue}{rgb}{0.53, 0.81, 0.92}
\definecolor{turquoise}{rgb}{0.19, 0.84, 0.78}
\definecolor{navyblue}{rgb}{0.0, 0.0, 0.5}
\definecolor{gray30}{rgb}{0.30078125,0.30078125,0.30078125}
\definecolor{gray80}{rgb}{0.796875,0.796875,0.796875}
\definecolor{slategray}{rgb}{0.44, 0.5, 0.56}
\definecolor{skyblue}{rgb}{0.53, 0.81, 0.92}
\definecolor{goldenrod}{rgb}{0.85, 0.65, 0.13}
\definecolor{brown(web)}{rgb}{0.65, 0.16, 0.16}
\definecolor{khaki(x11)(lightkhaki)}{rgb}{0.94, 0.9, 0.55}
\definecolor{beige}{rgb}{0.96, 0.96, 0.86}
\definecolor{purple(x11)}{rgb}{0.63, 0.36, 0.94}

\colorlet{mylinkcolor}{violet}
\colorlet{mycitecolor}{YellowOrange}
\colorlet{myurlcolor}{Aquamarine}

\newcommand{\nic}{\textcolor{Black}}

\begin{document}

\title{Can experimentally-accessible measures of entanglement distinguish 
quantum spin liquids from disorder--driven ``random singlet'' phases ?}

\author{Tokuro Shimokawa} 
\email{tokuro.shimokawa@oist.jp}
\affiliation{Theory of Quantum Matter Unit, Okinawa Institute of Science and Technology Graduate University, Onna-son, Okinawa 904-0412, Japan}

\author{Snigdh Sabharwal}
\affiliation{Theory of Quantum Matter Unit, Okinawa Institute of Science and Technology Graduate University, Onna-son, Okinawa 904-0412, Japan}

\author{Nic Shannon}
\affiliation{Theory of Quantum Matter Unit, Okinawa Institute of Science and Technology Graduate University, Onna-son, Okinawa 904-0412, Japan}

\date{\today}

\begin{abstract}

At the theoretical level, quantum spin liquids are distinguished 
from other phases of matter by their entanglement properties.  
However, since the usual measure of entanglement, 
entanglement entropy, cannot accessed in experiment, 
indentifying quantum spin liquids in candidate materials remains 
an acute problem.
Here we show other, experimentally--accessible, measures of entanglement 
can be used to distinguish a quantum spin liquid from a competing 
disorder--driven ``random singlet'' phase, in a model of a 
disordered antiferromagnet on a triangular lattice.  
The application of these results to the triangular--lattice systems 
YbZnGaO$_4$, YbZn$_2$GaO$_5$ and KYbSe$_2$ is discussed.

\end{abstract}

\maketitle


Half a century has passed since Anderson pioneered resonating valence bonds (RVB) 
as an alternative to low--temperature magnetic order \cite{Anderson1973}, paving the way 
for the modern concept of a quantum spin liquid (QSL) \cite{Balents2010}.   
In that time, the study of QSL has become a central topic in condensed matter 
physics~\cite{Lee2008,Savary2016,Zhou2017,Knolle2019}, 
with important connections to both quantum computing \cite{Kitaev2006-AnnPhys321}, 
and high-energy physics \cite{Yan2020}.
None the less, the identification of QSL in experiment remains a challenge, particularly in 
materials with chemical or structural disorder, where there can be many different routes 
to a spin--disordered ground state~\cite{Watanabe2014,Kawamura2014,Shimokawa2015,Uematsu2017,Wen2017,Zhu2017,Uematsu2018,Kimchi2018b,Liu2018b,Uematsu2019,Kawamura2019,Wu2019,Liu2020,Seki2021,Uematsu2021,Ren2023}.


In many cases, the sharpest distinction between conventional and unconventional phases of 
matter such as QSL lies in their entanglement properties \cite{Savary2016,Kitaev2006-PRL96,Levin2006,Laflorancie2016}.
This fact has been put to good use in simulation \cite{Laflorencie2005,Wu2019}, but when it comes 
to candidate materials, this approach is more challenging, since the entanglement 
entropy and entanglement spectra studied in simulation are not, in general, 
accessible to experiment \cite{Islam2015}.
None the less, there are alternative measures  or ``witnesses'' of entanglement which can 
be measured \cite{Brukner2006}.
Recent experiments have demonstrated the potential of this approach in context of molecular qubits \cite{Garlatti2017}, 
quasi--one dimensional spin chains \cite{Laurell2021,Scheie2021,Menon2023}, and 
triangular--lattice antiferromagents \cite{Pratt2022,Scheie2024,Wu-arXiv}.
However the signal properties of entanglement witnesses in spin models remain 
relatively unexplored, particularly when it comes to systems with disorder \cite{Sabharwal-arXiv}.
And an important question, in this context, is how we could experimentally 
distinguish a QSL from the 
\textcolor{black}{
disorder--driven phases found in systems 
with strongly disordered interactions 
\cite{Dasgupta1980,Fisher1994,Nonomura1995,Oitmaa2001,Motrunich2000,Lin2003,Igloi2005,Yu2006,Laflorencie2006,Tarzia2008,Vojta2009a,Vojta2009b,Singh2010,Thomas2011,Watanabe2014,Uematsu2017,Kimchi2018b,Igloi2018,Liu2018b,Kawamura2019,Liu2020,Sabharwal-arXiv}.
}


\begin{figure}[t!]
	\centering
	\subfloat[ Quantum spin liquid \label{fig:QSLCij} ]{\includegraphics[height=0.4\columnwidth]{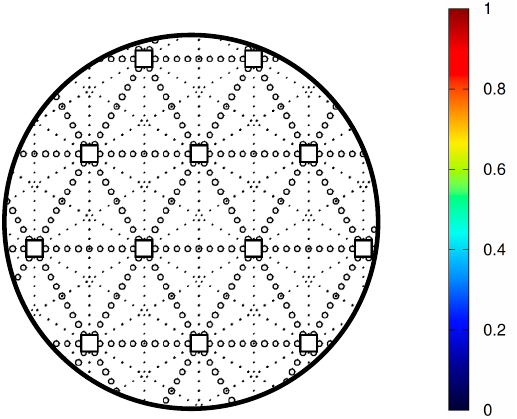}}
	\subfloat[ Random singlet \nic{phase} \label{fig:RSCij} ]{\includegraphics[height=0.4\columnwidth]{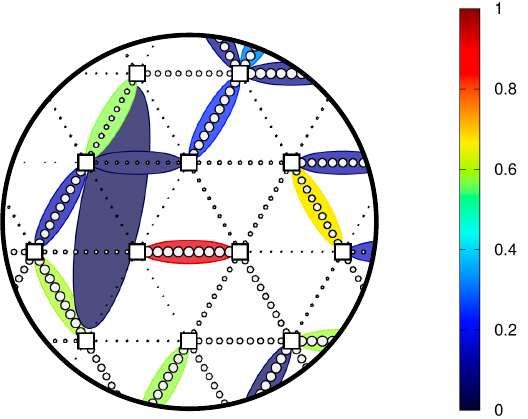}}
	\caption{\small  
		Singular difference in pairwise entanglement found in 
		quantum spin liquid (QSL) and disorder--driven ``random singlet'' (RS) phases 
		on a triangular lattice.
		(a) Concurrence $C_{ij}$ within QSL phase, showing that the entanglement 
		between any pair of spins is identically zero.
		(b) Equivalent results within RS phase, showing irregular pattern of entanglement 
		between pairs of spins on different lengthscales.
		Results are taken from zero--temperature exact diagonalization (ED) 
		calculations for a cluster of 30 spins, 
		for the disordered triangular--lattice antiferromagnet [Eq.~(\ref{eq:H})], with parameters 
		[Eq.~(\ref{eq:parameters})], as described in the text.
		The relative strength of interactions on different bonds is shown through
		thickness of 
		dashed lines. 
	}
	\label{fig:concurrence}
\end{figure}

\begin{figure}[t]
\includegraphics[width=7.5cm]{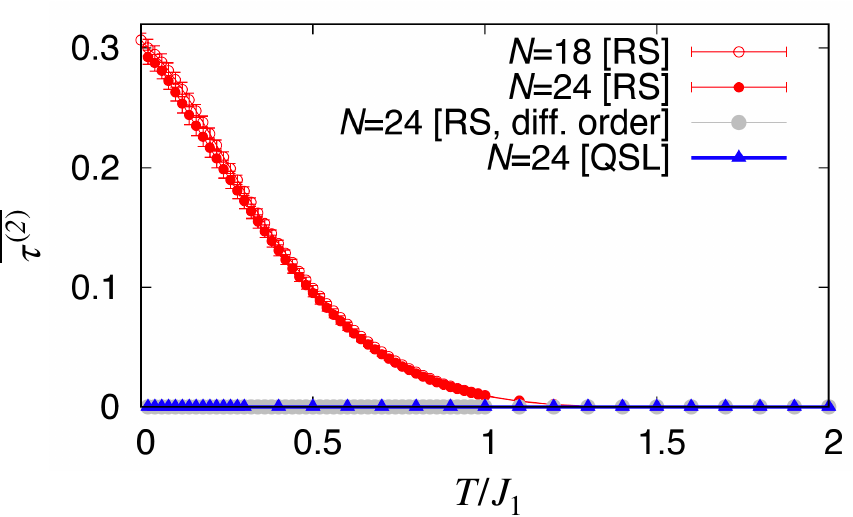}
\caption{
Temperature dependence of the two--tangle in the quantum spin liquid (QSL)
and disorder--driven random singlet (RS) phases. 
The two--tangle 
[Eq.~(\ref{eq:two.tangle.average})], 
averaged over different disorder realizations, 
takes on a finite value \mbox{$\overline\tau^{\sf (2)} > 0$} in the RS phase (red symbols), 
but vanishes exactly in the QSL (blue symbols), consistent with results for 
concurrence at \mbox{$T=0$}, shown in Fig.~\ref{fig:concurrence}.
However if concurrence $\tilde{C}_{ij}$ is calculated using the disorder--averaged
spin configurations measured in experiment [Eq~(\ref{eq:concurrence.in.experiment})], 
the resulting estimate of two--tangle also vanishes in the RS phase (grey symbols).
Results are taken from quantum typicality (TPQ) calculations for
[Eq.~(\ref{eq:H})], with parameters [Eq.~(\ref{eq:parameters})], 
as described in the text.
}
 \label{fig:two.tangle}
\end{figure}
\begin{figure*}
	\centering	
	\subfloat[ Temperature dependence of nQFI]{
  		\includegraphics[width=0.7\columnwidth]{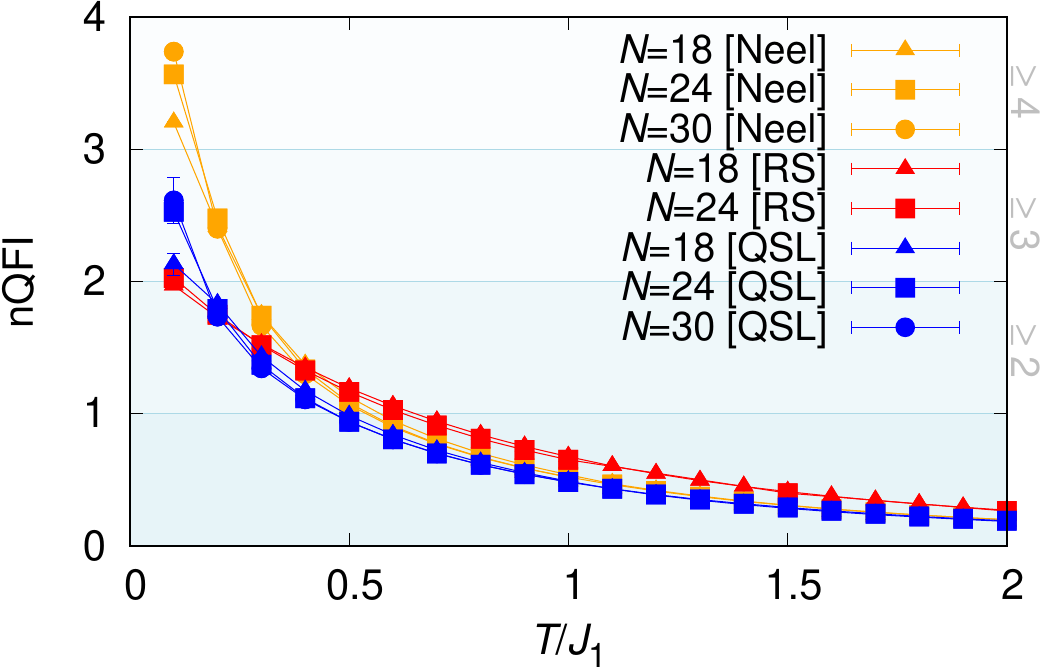}
		\label{fig:nQI.linear.scale}
	} 
	\subfloat[ nQFI in N\'eel and QSL \text{[log-log]}
	]{
  		\includegraphics[width=0.6\columnwidth]{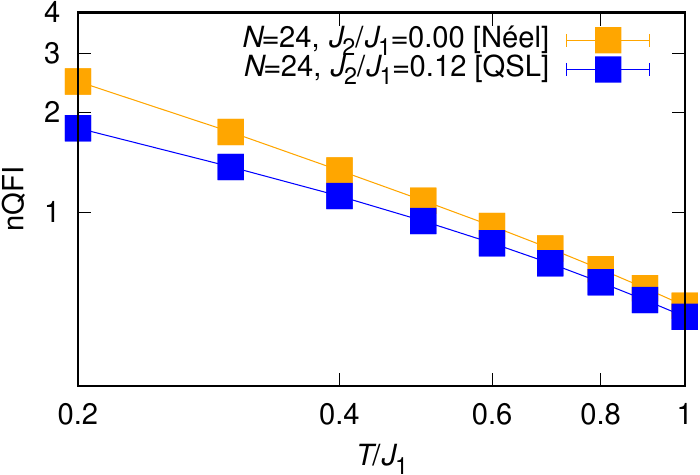}
		\label{fig:nQI.log.scale}
	}
	\subfloat[ nQFI in RS \text{[log-linear]}
	]{
  	\includegraphics[width=0.6\columnwidth]{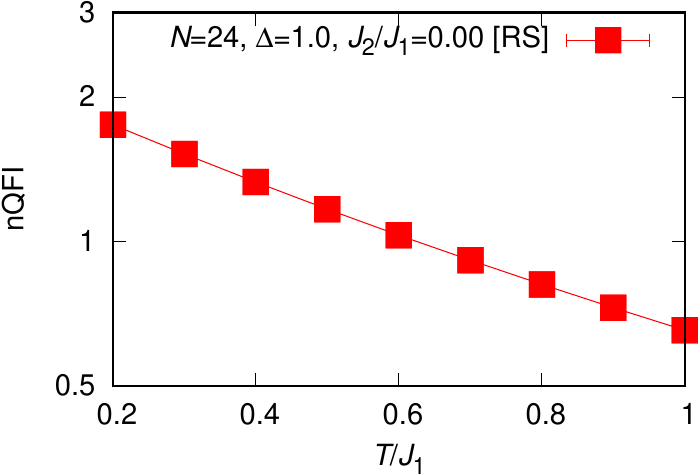}
	\label{fig:nQI.semi.log.scale}
	}
	\caption{
	Evolution of multipartite entanglement in N\'eel, QSL and RS phases 
	as a function of temperature, as witnessed by the quantum Fisher 
	information density (nQFI). 
	(a) Comparison of the temperature--dependence of nQFI [Eq.~(\ref{eq:nQFI})] 
	for the three different phases.
	In all cases, multipartite entanglement $m \geq 2$ is present at low temperature. 
	However results for the RS follow a different trajectory from those of 
	the N\'eel and QSL phases. 
	(b) Results for the N\'eel and QSL phases replotted on a log--log scale, 
	showing the power--law growth of the depth of entanglement as $T \to 0$ 
	[Eq.~(\ref{eq:QFI.power.law})].  
	(c) Results for the RS phase replotted on a semi--log scale, showing  
	exponential suppression of the depth of entanglement with increasing 
	temperature [Eq.~(\ref{eq:QFI.exponential})].
	Results are taken from quantum typicality (TPQ) calculations for
	[Eq.~(\ref{eq:H})], with parameters [Eq.~(\ref{eq:parameters})], 
	as described in the text.
	In all cases, results for nQFI were evaluated at the N\'eel ordering vector  
 	 \mbox{${\bf q}_{\sf K}$} [Eq.~(\ref{eqn:qK})]. 
	}
	\label{fig:QFI}
\end{figure*}
\begin{figure*}
\centering
	\begin{minipage}{0.3\textwidth}
		\centering
		\subfloat[nQFI found in experiment and simulation \label{fig:nQFI.experiment}]{
  			\includegraphics[width=\textwidth]{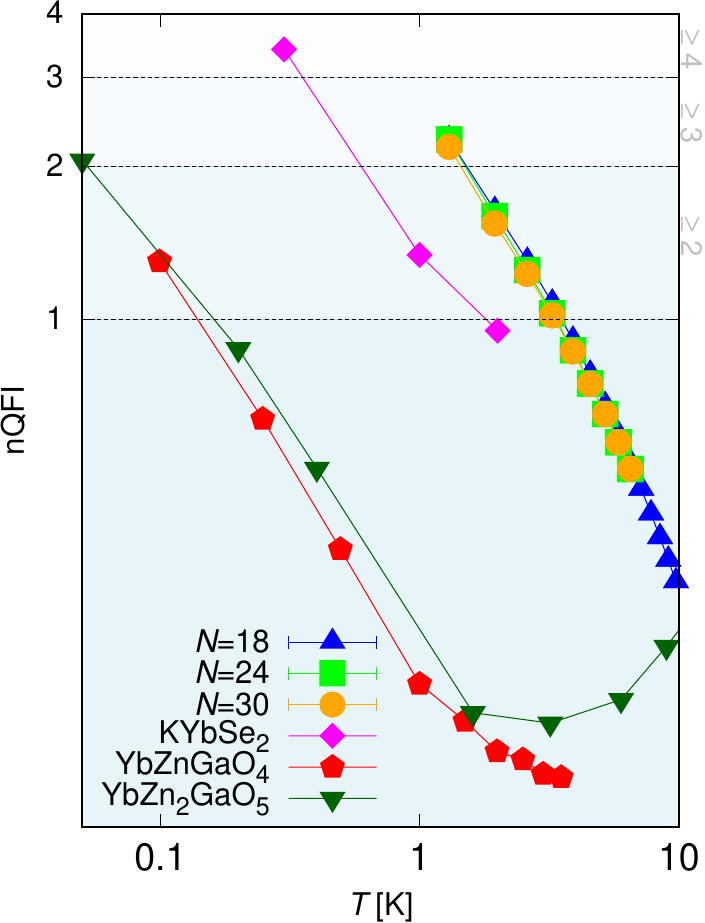}
		}
	\end{minipage}
	\hspace{0.5cm}
	\begin{minipage}{0.50\textwidth}
		\centering
		\subfloat[S(${\bf q}_{\sf K}$, $\omega$, T) \text{[KYbSe$_2$]} ]{
 			\includegraphics[width=0.50\textwidth]{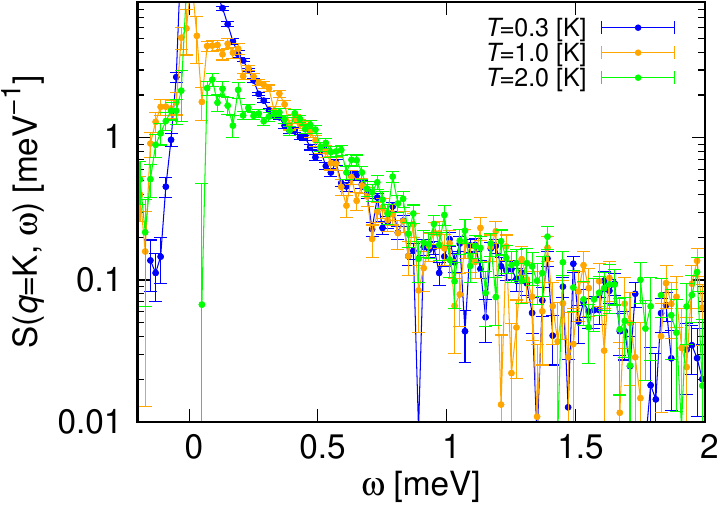}
			\label{fig:SqwT.KYbSe2}
		}
		\subfloat[Scaling collapse  \text{[KYbSe$_2$]} ]{
 			\includegraphics[width=0.52\textwidth]{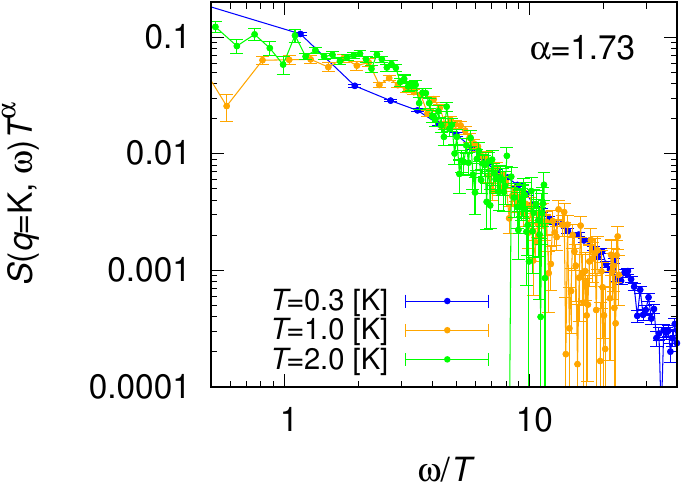}
			\label{fig:scaling.collpase.KYbSe2}
                } \\
		\subfloat[S(${\bf q}_{\sf K}$, $\omega$, T) \text{[simulation]} ]{
  			\includegraphics[width=0.50\textwidth]{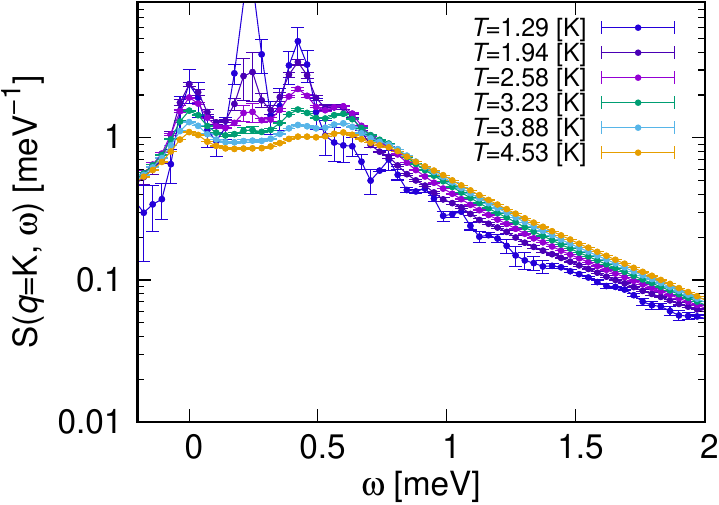}
			\label{fig:SqwT.simulation}	}
		\subfloat[Scaling collapse \text{[simulation]} ]{
 			\includegraphics[width=0.52\textwidth]{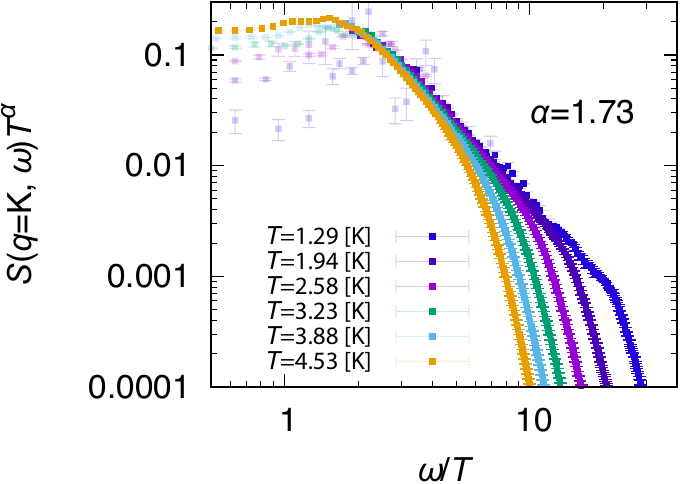}
			\label{fig:scaling.collpase.simulation}}
	\end{minipage}
	\caption{
	Comparison of results for quantum Fisher information density, and 
	\nic{scaling collapse of dynamical susceptibility, as} 
	found in simulation and experiment.    
	(a) Quantum Fisher information density (nQFI) 
	as found in simulation, and experiments 
	on KYbSe$_2$ \cite{Scheie2024}, YbZnGaO$_4$ \cite{Pratt2022}
	and YbZn$_2$GaO$_5$ \cite{Wu-arXiv}.
	Both theory and experiment are consistent with a depth of entanglement 
	which diverges as a power law for $T \to 0$, characteristic of 
	quantum spin liquid (QSL) and N\'eel phases.   
	(b) Evolution of dynamical structure factor of KYbSe$_2$ as function 
	of temperature \cite{Scheie2024}.
	(c) Scaling collapse of experimental data as a function of $\omega/T$, 
	consistent with the form expected near a quantum critical point (QCP).   
	(d) Evolution of dynamical structure factor found in simulation as function 
	of temperature.
	(e) Scaling collapse of simulation data.
	Simulation results are taken from quantum typicality (TPQ) calculations for
	[Eq.~(\ref{eq:H})], with parameters [Eq.~(\ref{eq:parameters.KYbSe2})], 
	as described in the text.
	In all cases, 
	fluctuations were evaluated at the N\'eel ordering vector
	\mbox{${\bf q}_{\sf K}$} [Eq.~(\ref{eqn:qK})].
	}
	\label{fig:experiment}
\end{figure*}

In this Letter, we show that experimentally-accessible measures of entanglement 
can distinguish a quantum spin liquid from the disorder--driven ``random singlet'' 
phase in a representative model of a triangular lattice antiferromagnet.
The measures we consider are 
concurrence, and the closely related two--tangle~\cite{Coffman2000,Amico2004,Amico2006}, 
and quantum Fisher information~\cite{Pezze2009,Helmut2014,Hauke2016}, 
both of which have previously been discussed in the context of inelastic neutron scattering 
or muon spin relaxation experiments~\cite{Brukner2006,Garlatti2017,Laurell2021,Scheie2021,Pratt2022,Scheie2024,Laurell2024,Scheie2025,Wu-arXiv}.  
These are calculated at finite temperature, within a quantum 
typicality approach derived from exact diagonalization~\cite{Imada1986,Ham2000,Sugiura2013}. 
We find that both RS and QSL phases exhibit multipartite entanglement, but show very different 
characteristics in pairwise entanglement, as measured through concurrence [Fig.~\ref{fig:concurrence}].
While the temperature--dependence of the two-tangle (or concurrence) clearly 
distinguishes RS from QSL \nic{phases} in simulation, we conclude that these 
measures of entanglement cannot be safely inferred from experiments on disordered systems 
[Fig.~\ref{fig:two.tangle}].
In contrast, we find that the temperature--dependence of quantum Fisher information (QFI) 
provides a robust tool for distinguishing RS from QSL states 
\nic{in triangular--lattice antiferromagnets} [Fig.~\ref{fig:QFI}].
We discuss the implication of these results for experiment, 
finding evidence in support of spin liquid 
scenario for YbZnGaO$_4$~\cite{Pratt2022} and YbZn$_2$GaO$_5$~\cite{Wu-arXiv}, 
and showing that and the properties of KYbSe$_2$~\cite{Scheie2024} are consistent 
with proximity to a quantum critical point [Fig.~\ref{fig:experiment}].

{\it Model and materials}.\ 
%
The competition between N\'eel, QSL and disorder--driven 
``random singlet'' phases has been widely discussed in the context 
of triangular--lattice antiferromagnets (AF) based on (pseudo)spin--1/2 
Yb$^{3+}$ ions \cite{Liu2018a,Schmidt2021}, including 
NaYbS$_2$ \cite{Baenitz2018,Sarkar2019}, 
NaYbO$_2$ \cite{Ding2019,Bordelon2019}, 
NaYbSe$_2$ \cite{Ranjith2019,Dai2021},
RbYbSe$_2$ \cite{Xing2021}, 
CsYbSe$_2$ \cite{Xie2023} and 
YbZn$_2$GaO$_5$ \cite{Wu-arXiv}.
A prototypical example is YbMgGaO$_4$, originally proposed 
as a QSL  \cite{Li2015a,Li2015b,Shen2016,Li2016,Paddison2017,Li2017,Iaconis2018,Ding2020,Majumder2020}, which has since been 
argued to exhibit a disorder--driven state 
\textcolor{black}{
mimicing the correlations 
of a QSL \cite{Zhu2017,Kimchi2018b,Parker2018}.  
}
Isostructural YbZnGaO$_4$ has been discussed as both 
a QSL \cite{Pratt2022,Wu-arXiv} and a spin glass \cite{Ma2018}.
Meanwhile KYbSe$_2$ exhibits N\'eel order,  but has been argued 
to lie close to a quantum critical point (QCP) separating this 
from a QSL \cite{Scheie2021}.


With these results in mind, we consider the model most commonly used to interpret 
experiments on these materials, 
 \begin{eqnarray}
    \mathcal{H} 
   = J_1 \sum_{\langle ij \rangle_1} (1+ \Delta \alpha_{ij}) {\bf S}_i \cdot {\bf S}_j 
   + J_2 \sum_{\langle ij \rangle_2} (1+ \Delta \beta_{ij}) {\bf S}_i \cdot {\bf S}_j  \; , 
    \nonumber  \\
    \label{eq:H}
 \end{eqnarray}
where interactions on 1$^{st}$-- and 2$^{nd}$--neighbour bonds 
of a triangular lattice are subject to quenched random disorder of strength 
$\Delta$, drawn from a uniform distribution   
 \begin{eqnarray}
    \alpha_{ij} , \beta_{ij} \in [-1, 1] \; .
 \end{eqnarray}

\begin{figure}[t]
	\centering
	  \includegraphics[width=8.0cm]{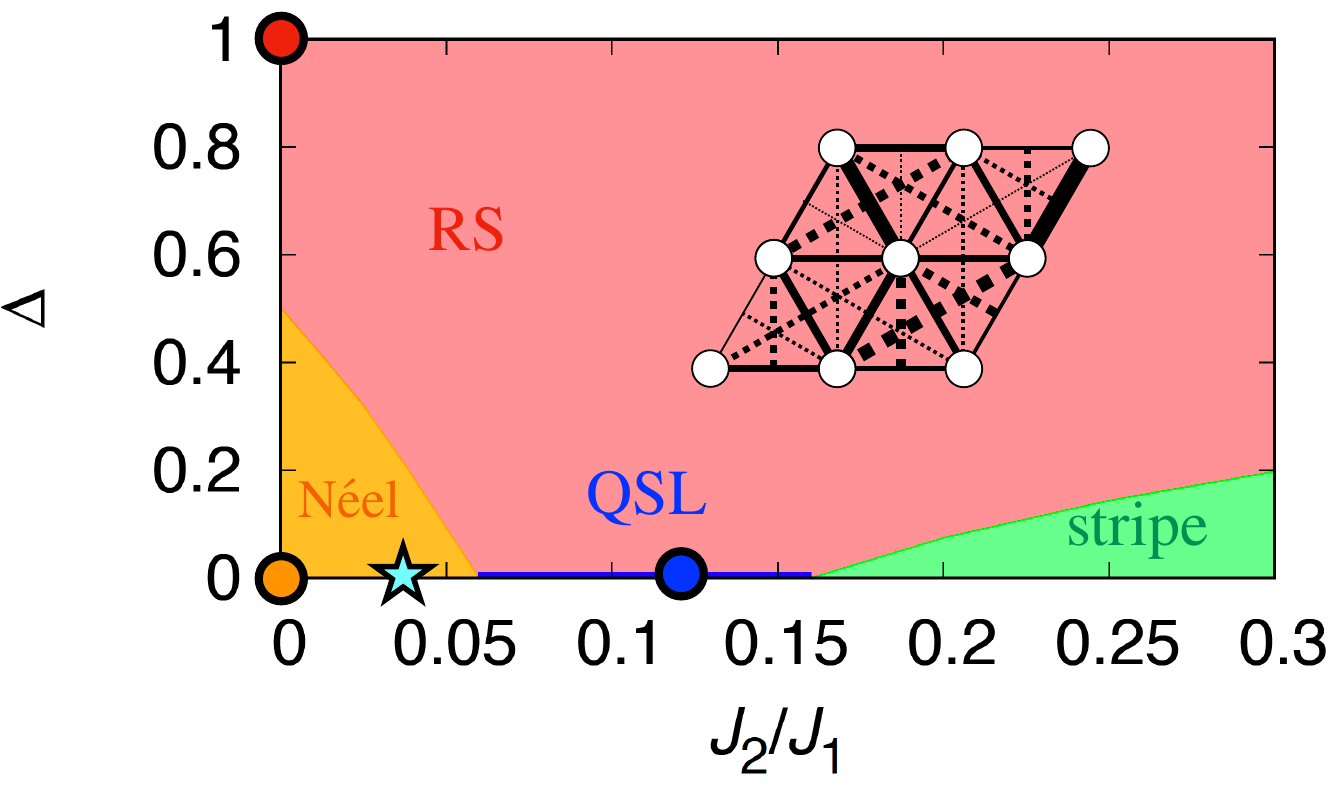} 
	  \label{fig:model}
	\caption{\small  
	Schematic ground state phase diagram for the spin--1/2 $J_1$--$J_2$ 
	antiferromagnet on a triangular lattice, as a function of bond--disorder.
	In the absence of disorder \mbox{($\Delta = 0$)}, ordered antiferromagnetic 
	``N\'eel'' and ``stripe'' phases are separated by a region where the ground 
	state is identified as a quantum spin liquid (QSL).
	In the presence of disorder \mbox{($\Delta > 0$)}, these phases give 
	way to a disorder--driven state which can mimic many of the properties of the QSL, 
	denoted here as ``random singlet'' (RS) phase.   
	Phase boundaries reflect the general trends in 
	published results for the model [Eq.~(\ref{eq:H})],  
	in the presence \cite{Watanabe2014,Wu2019}
	and absence \cite{Kaneko2014,Li2015b,Zhu2015,Hu2015}
	of disorder. 
	The inset illustrates interactions $J_1$ 
	(solid lines) and $J_2$ 
	(dashed lines) at finite $\Delta$, with the thickness 
	of the line reflecting the strength of the interaction.
	Parameter sets considered in this paper [Eq.~(\ref{eq:parameters})]
	are shown with circles.   		
	A star denotes parameters proposed for  
	KYbSe$_2$ \cite{Scheie2024}.
	}
	\label{fig:phase.diagram}
\end{figure}

In Fig.~\ref{fig:phase.diagram} we show a schematic phase diagram of the 
model Eq.~(\ref{eq:H}), based on previous studies with \cite{Watanabe2014,Shimokawa2015,Wu2019}, 
and without 
\cite{Kaneko2014,Li2015b,Zhu2015,Hu2015,Iqbal2016,Saadatmand2016,Gong2019,Hu2019,Jiang2023}, 
disorder.
Interactions are shown schematically in an inset.
In the absence of disorder ($\Delta =0$), the model supports \mbox{3--sublattice} coplanar N\'eel order 
for small $J_2/J_1$ 
\cite{Bernu1994}, and \mbox{2--sublattice} collinear ``stripe'' order 
for large $J_2/J_1$ 
\cite{Jolicoeur1990,Chubukov1992}.
A quantum spin liquid (QSL), which may be gapless \cite{Kaneko2014, Gong2019, Hu2019,Wietek2024}, 
or posses a small gap  \cite{Zhu2015, Saadatmand2016,Hu2015,Jiang2023},  
has been identified at intermediate parameters \footnote{\protect{For a discussion of possible valence bond state, see \cite{Wietek2024}}.}.  
In the presence of disorder $\Delta > 0$, both QSL and ordered phases give way 
to a disorder--driven state which can mimic many of the properties of the 
QSL \cite{Watanabe2014,Wu2019}, here labelled ``RS''.


In one dimension (1D), bond--disorder of the type found in Eq.~(\ref{eq:H}) gives rise to a 
``random--singlet'' (RS) phase whose long--distance properties are controlled by singlets 
formed at exponentially--long length scales \cite{Dasgupta1980,Fisher1994}.
The RS phase mimics a number of the properties of the critical Luttinger liquid (1D QSL) 
found in spin chains without disorder, including an algebraic $1/r^2$ decay of correlations,  
and logarithmic growth of entanglement entropy $S_{\sf E} \sim \log L$ \cite{Refael2004,Laflorencie2005,Sabharwal-arXiv}.
None the less it represents an entirely distinct phase of matter, 
lacking the conformal invariance and coherent fractionalized excitations (spinons) 
of the Luttinger liquid, and instead having properties controlled by the fixed point of a 
strong--disorder renomalization group (SDRG) \cite{Fisher1994}.


The situation for disordered magnets in two dimensions (2D) is much less clear. 
Early arguments of Imry and Ma suggest that disorder can have singular 
effect on gapless ordered states in 2D \cite{Imry1975}, 
while SDRG calculations identify a potential fixed point 
with spin glass character \cite{Lin2003,Igloi2005,Igloi2018},  
\textcolor{black}{
and a number of other scenarios have also been discussed 
\cite{Nonomura1995,Motrunich2000,Oitmaa2001,Lin2003,Yu2006,Laflorencie2006,Yu2006,Tarzia2008,Vojta2009a,Vojta2009b,Singh2010,Thomas2011,Kimchi2018b}
}.
Meanwhile numerical calculations for frustrated magnets with bond disorder 
\cite{Kawamura2014,Watanabe2014,Shimokawa2015,Zhu2017,Uematsu2017,Liu2018b,Uematsu2018,Uematsu2017,Kawamura2019,Wu2019,Uematsu2019,Liu2020,Uematsu2021}
find disorder--driven states which can mimic both the thermodynamic properties 
of gapless QSL, \mbox{$C(T\to 0) \propto T^\alpha$} \cite{Kawamura2014}, 
and the diffuse structures found in the correlations $S({\bf q})$ \cite{Kawamura2014,Zhu2017,Kimchi2018b}.   
Results for different models display a number of common features, 
including the presence of ``random'' (static) singlet bonds over multiple length 
scales and the existence of orphaned spins \cite{Kawamura2014,Kimchi2018b}.   
None the less, the general classification of numerical results in terms of RG 
fixed points 
remains an open problem.   
Here, we follow \cite{Kawamura2014,Watanabe2014,Shimokawa2015,Liu2018b,Liu2020} in using the term 
``random singlet'' (RS) phase to describe the disorder--driven phase we study, 
without seeking to identify this with any particular fixed point.  


For purposes of numerical study, we select parameters characteristic of the 
N\'eel, QSL and RS phases 
\begin{subequations}
 \begin{eqnarray}
	\textcolor{orange}{\CIRCLE} \qquad (J_1, J_2, \Delta) &=& (1,0,0)  \qquad  [\text{N\'eel}]  \\
	\textcolor{blue}{\CIRCLE} \qquad (J_1, J_2, \Delta) &=& (1,0.12,0)  \quad [\text{QSL}]   \\
	\textcolor{red}{\CIRCLE} \qquad (J_1, J_2, \Delta) &=& (1,0,1) \qquad   [\text{RS}] 	 
\end{eqnarray}
\label{eq:parameters}
\end{subequations}
as shown in Fig.~\ref{fig:phase.diagram}.
We also consider one parameter set taken from experiments on KYbSe$_2$ \cite{Scheie2024,Scheie2024-PRB109} 
\begin{eqnarray}
\textcolor{cyan}{\star} \qquad (J_1, J_2, \Delta) 
	&=& (1,0.047,0)  \qquad  [\text{KYbSe$_2$}] 
	\label{eq:parameters.KYbSe2}
\end{eqnarray}
Numerical simulations of Eq.~(\ref{eq:H}) were carried out for clusters of 
\mbox{$N = 18, 24, 30$} spins, using exact diagonalization (ED) at $T=0$
and thermal pure quantum state (TPQ) methods at finite $T$, as described 
in the Supplemental Material \cite{supplemental-material}.

{\it Entanglement measures.}\  
%
We consider three different measures of entanglement  
concurrence, 
two--tangle, 
and quantum Fisher information.   
Since these measures apply to mixed states, and can be expressed 
in terms of two--point spin correlation functions,  
they are, \nic{{\it a priori},  directly applicable} 
to experiment~\cite{Brukner2006,Garlatti2017,Laurell2021,Pratt2022,Scheie2024}. 
The definitions given 
below apply to systems 
model with (at least) $U(1)$ spin--rotation symmetry; further details 
can be found in two recent reviews \cite{Laurell2024,Scheie2025}.  


{\it Concurrence:}
   \begin{eqnarray}
     C_{ij} &=& 2\ {\rm max} \Large\{ 0, \sqrt{(\langle S_i^xS_j^x \rangle + \langle S_i^y S_j^y \rangle)^2}  \nonumber \\
     &-&  \sqrt{\left(\frac{1}{4}+\langle S_i^z S_j^z  \rangle \right)^2
     -\left(\frac{\langle S_i^z \rangle + \langle S_j^z \rangle}{2}\right)^2}    \Large\} \; , 
     \label{eq:concurrence}
   \end{eqnarray}
provides a quantitative measure of the entanglement between a given pair of spins, 
with $C_{ij} \in [0,1]$ where  $0$ signals a separable state, and $1$ maximal entanglement~\cite{Coffman2000,Amico2004,Amico2006}.  


{\it Two-tangle}:
    \begin{eqnarray}
      \tau^{(2)}_j =  \sum_{i \ne j} C_{ij}^2 \; , 
      \label{eq:two.tangle}
   \end{eqnarray} 
is derived from concurrence, and quantifies the total two--spin entanglement associated 
with a spin on site $j$~\cite{Coffman2000,Amico2004,Amico2006}.  


{\it Quantum Fisher information density (nQFI)}:
\textcolor{black}{
   \begin{eqnarray}
     f_{\mathcal{Q}} [{\bf q},T] &=& 4 \int_{0}^{\infty}  d\omega \  {\rm tanh}
     \left( \frac{\omega}{2 T} \right)(1-{\rm e}^{-\beta \omega}) S^{zz} ({\bf q}, \omega, T) , \nonumber \\
     \label{eq:nQFI} 
    \end{eqnarray} 
}
\nic{is defined in terms of a dynamical structure factor (here, $S^{zz} ({\bf q}, \omega, T)$),  
and provides a lower bound on the depth of entanglement 
associated with fluctuations at a given wavevector~${\bf q}$.}
More concretely, when $f[{\bf q}, T]>m$ (where $m$ is a positive integer  
and a divisor of $N$), the quantum state \nic{in question} is at least $m+1$ multipartite 
entangled~\cite{Pezze2009,Hyllus2012,Toth2012,Helmut2014,Hauke2016}.


{\it Concurrence as measured by experiment}:
\begin{eqnarray}
     \tilde{C}_{ij} &=& 2\ {\rm max} \{0, \sqrt{(\overline{\langle S_i^xS_j^x \rangle} + \overline{\langle S_i^y S_j^y \rangle})^2}  \nonumber \\
     &-&\sqrt{\left(\frac{1}{4}+\overline{\langle S_i^z S_j^z  \rangle}\right)^2
     - \left(\frac{\overline{\langle S_i^z \rangle} + \overline{\langle S_j^z \rangle}}{2}\right)^2}    \}.
     \label{eq:concurrence.in.experiment}
\end{eqnarray}
In theoretical calculations, these measures of entanglement can straightforwardly 
be generalized to systems with disorder by taking  
averages of Eq.~(\ref{eq:concurrence}),  Eq.~(\ref{eq:two.tangle}) and 
 Eq.~(\ref{eq:nQFI}) within an ensemble of disordered states.
We denote these as quantities as $\overline{C_{ij}}$, $\overline{\tau^{(2)}}$ and $\overline{f_{\mathcal{Q}}}$, respectively.    
However when evaluating entanglement measures from experiment, 
it is instead the disorder--averaged correlation of spins 
\mbox{$\overline{\langle S_i^\alpha S_j^\alpha \rangle}$} 
which enters calculations, leading to Eq.~(\ref{eq:concurrence.in.experiment}).
This difference in the order of averages has profound implications for the 
values of concurrence in the presence of disorder, a point discussed 
at length in \cite{Sabharwal-arXiv}.


{\it Results for concurrence.}\  
%
The singular difference between the entanglement structure of the QSL and 
RS phases is evident at the level of the two--spin entanglement probed 
by concurrence.  
In Fig.~\ref{fig:concurrence} we show results for $C_{ij}$ [Eq.~(\ref{eq:concurrence})]
in  QSL and RS ground states, where results for the RS are shown for a single 
realization of disorder.
Concurrence within the QSL vanishes identically, reflecting the fact that 
its entanglement is of a profoundly multipartite nature.    
The RS phase also possess multipartite entanglement, 
as evidenced by the residual tangle, discussed in \cite{supplemental-material}, 
and the QFI, discussed below.
None the less, individual spins \nic{retain} the freedom to form singlets with other spins.
These singlets occur at different length scales, with singlets on 
1$^{st}$--neighbour bonds only weakly correlated with the strength 
of interactions on that bond.


{\it Results for two tangle.}\   
%
\nic{
At high temperatures, the (deeply) entangled QSL and RS ground states must 
ultimately give way to a trivial paramagnetic phase, without entanglement.
}
In Fig.~\ref{fig:two.tangle}, we compare results for the temperature dependence of 
the two--tangle 
\begin{eqnarray}
	\tau^{(2)} = \frac{1}{N} \sum_i \tau^{(2)}_i
\label{eq:two.tangle.average}
\end{eqnarray}
%
in the QSL and RS phases, 
where results for the RS have been averaged over multiple realizations of disorder.  
For this order of averages, we recover the behavior expected from ground state 
correlations, with results for the QSL vanishing identically, and the two--spin 
entanglement of RS saturating at at finite value for \mbox{$T \to 0$}.   
However, if the order of averages is reversed, to reflect the correlations measured in 
experiment [Eq.~(\ref{eq:concurrence.in.experiment})], results for the two--tangle 
also vanish in the RS phase. 


The quenching of the two--tangle by disorder is a direct consequence 
of the convex roof function used in the definition of concurrence \cite{Sabharwal-arXiv}.
It follows that experimental measurements systematically underestimate the 
value of concurrence, relative to the Eq.~(\ref{eq:concurrence}), and must 
be approached with extreme caution.
%
%
For this reason, in what follows below, we pursue other 
measures which are reliably accessible in experiment.
None the less, concurrence, two tangle 
and the closely related one--tangle \cite{Coffman2000,Amico2004,Amico2006}, 
remain useful tools for interpreting simulations results.
As a practical exmple, in  \cite{supplemental-material}, we show how 
these measures can be used to determine the phase
boundary between N\'eel and RS phases found 
in diagonalization calculations for 
Eq.~(\ref{eq:H}) [cf. Fig.~\ref{fig:phase.diagram}].

{\it Results for quantum Fisher information.}\  
%
While concurrence may prove an unreliable witness in disordered
materials, no such problems arise in evaluating quantum Fisher information.  
%
In Fig.~\ref{fig:QFI} we present results for the quantum Fisher information density, 
$f_{\mathcal{Q}}[{\bf q}, T]$ [Eq.~(\ref{eq:nQFI})], 
for parameters representative of N\'eel, QSL and RS phases [Eq.~(\ref{eq:parameters})].
In all three phases, the dominant two--spin correlations are found at 
\begin{eqnarray}
	{\bf q}_{\sf K} = \left(\frac{4\pi}{3},0\right)
\label{eqn:qK}
\end{eqnarray}
\cite{supplemental-material}, and we therefore use this wavevector
in calculating quantum Fisher information.   
In all cases, the depth of entanglement grows at low temperatures, with all 
three phases showing (at least) 3--partite entanglement at the lowest temperatures 
considered [Fig.~\ref{fig:nQI.linear.scale}].  
However, while results suggest that the depth of entanglement in both the N\'eel and QSL 
phases diverges as $T\to 0$, the temperature dependence in RS is qualitatively 
different, and consistent with saturation to a finite value for $T\to 0$.


We find that data for N\'eel and QSL are consistent with a power law 
divergence at low temperatures [Fig.~\ref{fig:nQI.log.scale}], 
 \begin{eqnarray}
    f_{\mathcal{Q}}[{\bf q}_{\sf K}, T/J_1] 
    	&\propto& b (T/J_1)^{-c}  \; \text{[N\'eel, QSL]} \; , 
\label{eq:QFI.power.law} 
\end{eqnarray}
while data for the RS exhibit exponential decay with increasing temperature [Fig.~\ref{fig:nQI.semi.log.scale}],  
 \begin{eqnarray}
    f_{\mathcal{Q}}[{\bf q}_{\sf K}, T/J_1] 
    	&\propto& b' e^{-c' (T/J_1)}   \; \text{[RS]} \; .
\label{eq:QFI.exponential} 
\end{eqnarray}
The distinction between power law and exponential decay of QFI 
as function of temperature marks a sharp, qualitative, difference 
QSL and RS which could be used to distinguish them in experiment.


{\it Comparison with experiment.}\  
%
QFI has been measured in three different triangular--lattice AF's:  
YbZnGaO$_4$~\cite{Pratt2022} and YbZn$_2$GaO$_5$~\cite{Wu-arXiv}, 
through muon spin rotation ($\mu$SR);  
and KYbSe$_2$, through inelastic neutron scattering (INS) \cite{Scheie2021}.
In Fig.~\ref{fig:nQFI.experiment} we show results for the scaling of the 
QFI density found in these experiments,  
and in simulations for the parameter set proposed for KYbSe$_2$ [Eq.~(\ref{eq:parameters.KYbSe2})].
In all three cases we find evidence for a power--law divergence 
in nQFI at low temperature [Eq.~(\ref{eq:QFI.power.law})], 
consistent with simulation results for N\'eel and QSL phases 
[Fig.~\ref{fig:nQI.log.scale}].
We draw two conclusions from these results:
\begin{enumerate}

\item The temperature--dependence of nQFI measured 
in KYbSe$_2$  
is consistent with the observed N\'eel order.

\item The temperature--dependence of nQFI measured 
in YbZnGaO$_4$ and YbZn$_2$GaO$_5$ 
distinguishes in favour of a QSL scenario, and against a RS scenario
for these materials.

\end{enumerate}
%
{\it Connections with quantum criticality.}\   
%
The power--law scaling seen in nQFI [Eq.~(\ref{eq:QFI.power.law})], 
is reminiscent of critical phenomena in the limit where $T_c \to 0$, i.e. 
approaching a quantum critical point (QCP) \cite{Chakravarty1989,Chubukov1994a,Chubukov1994b}.
In \cite{Scheie2024}, it was found that the dynamical structure factor 
measured in KYbSe$_2$ [Fig.~\ref{fig:SqwT.KYbSe2}] collapses onto 
a single function 
\begin{eqnarray}
	T^\alpha S({\bf q}_{\sf K}, \omega, T) = \Phi \left( \frac{\hbar \omega}{k_B T} \right) \; ,
\label{eq:scaling.collapse}
\end{eqnarray}
[Fig.~\ref{fig:scaling.collpase.KYbSe2}], a result which was interpreted 
in terms of proximity to a QCP.
In Fig.~\ref{fig:SqwT.simulation} we show simulation results for 
$S({\bf q}_{\sf K}, \omega, T)$, for a cluster of \mbox{$N=30$} spins 
with parameters \nic{taken from experiments on} KYbSe$_2$ [Eq~(\ref{eq:parameters.KYbSe2})].  
%
%
Simulation results also a exhibit a scaling collapse of the 
form Eq.~(\ref{eq:scaling.collapse}), over a range of $\omega/T$ determined
by the size of cluster (scatter of points at small $\omega/T$), 
and the timestep used in simulations, (suppression of response  
at large $\omega/T$) [Fig.~\ref{fig:scaling.collpase.simulation}].
\nic{A good collapse can be obtained using the exponent reported from 
experiment \mbox{$\alpha =1.73$}~\cite{Scheie2024}.
These results lends support to the scenario that the properties of KYbSe$_2$ 
are influenced by a nearby QCP~\footnote{Further analysis of the exponent $\alpha$ 
required to collapse data is provided in \cite{supplemental-material}}.}

{\it Conclusions.}\  
%
In this Letter we have addressed the question of whether experimentally--accessible 
measures of entanglement can distinguish quantum spin liquids (QSL) from disorder--driven 
``random singlet'' (RS) phases.
We have addressed this question in the context of model  
motivated by  Yb--based triangular lattice antiferromagnets, 
which have been proposed as QSL candidates, and where 
the effects of disorder have also been called into question.
And in this \nic{context}, we find that the answer is ``{\it yes}'': 
the way in which the depth of entanglement grows as temperature is 
reduced, witnessed by the quantum Fisher information (QFI), sharply 
distinguishes QSL [Fig.~\ref{fig:nQI.log.scale}] 
from RS [Fig.~\ref{fig:nQI.semi.log.scale}] phases.

These results can be compared directly with measurements of QFI in 
triangular lattice antiferromagnets YbZnGaO$_4$~\cite{Pratt2022} 
and YbZn$_2$GaO$_5$~\cite{Wu-arXiv}, where they provide evidence 
in support of a QSL [Fig.~\ref{fig:nQFI.experiment}].
We also revisit experiments on KYbSe$_2$ \cite{Scheie2024}, 
finding results consistent with N\'eel order, proximate to a quantum 
critical point [Fig.~\ref{fig:experiment}].
These findings confirm the utility of  measures of entanglement 
as a practical tool for distinguishing QSL phases in experiment.  

${\it Acknowledgments.}$----
%
This work was supported by the Theory of Quantum Matter Unit of the 
Okinawa Institute of Science and Technology Graduate University (OIST), 
and is also supported by JSPS Grant-in-Aid for Scientific Research (C) 
Grants No. 21K03477, 25K07213 and MEXT Grant-in-Aid for Transformative 
Research Areas A {\it ``Extreme Universe''} Grant No. 22H05266 and 24H00974.
The authors are pleased to acknowledge helpful conversations with 
\nic{Matthias Gohlke, Hikaru Kawamura, Hitoshi Ohta,  Pranay Patil, 
Geet Rakala, and Anders Sandvik}, 
and are grateful to Allen Scheie, Alan Tennant and Francis Pratt 
for providing the initial \nic{inspiration} for this research.
%
Numerical calculations were carried out using HPC facilities provided by  
the Supercomputing Center, ISSP, the University of Tokyo; OIST; and the 
supercomputer Fugaku, provided by RIKEN through the 
HPCI System Research project (Project IDs:hp230114, hp240014 and hp250047).   
T.~S. thanks Hiroki Nakano for assistance in using Fugaku.



\bibliography{paper.bib}

\onecolumngrid
\begin{center}
    \includegraphics[page=1,scale=1.0, trim=18mm 0mm 0mm 10mm, clip]{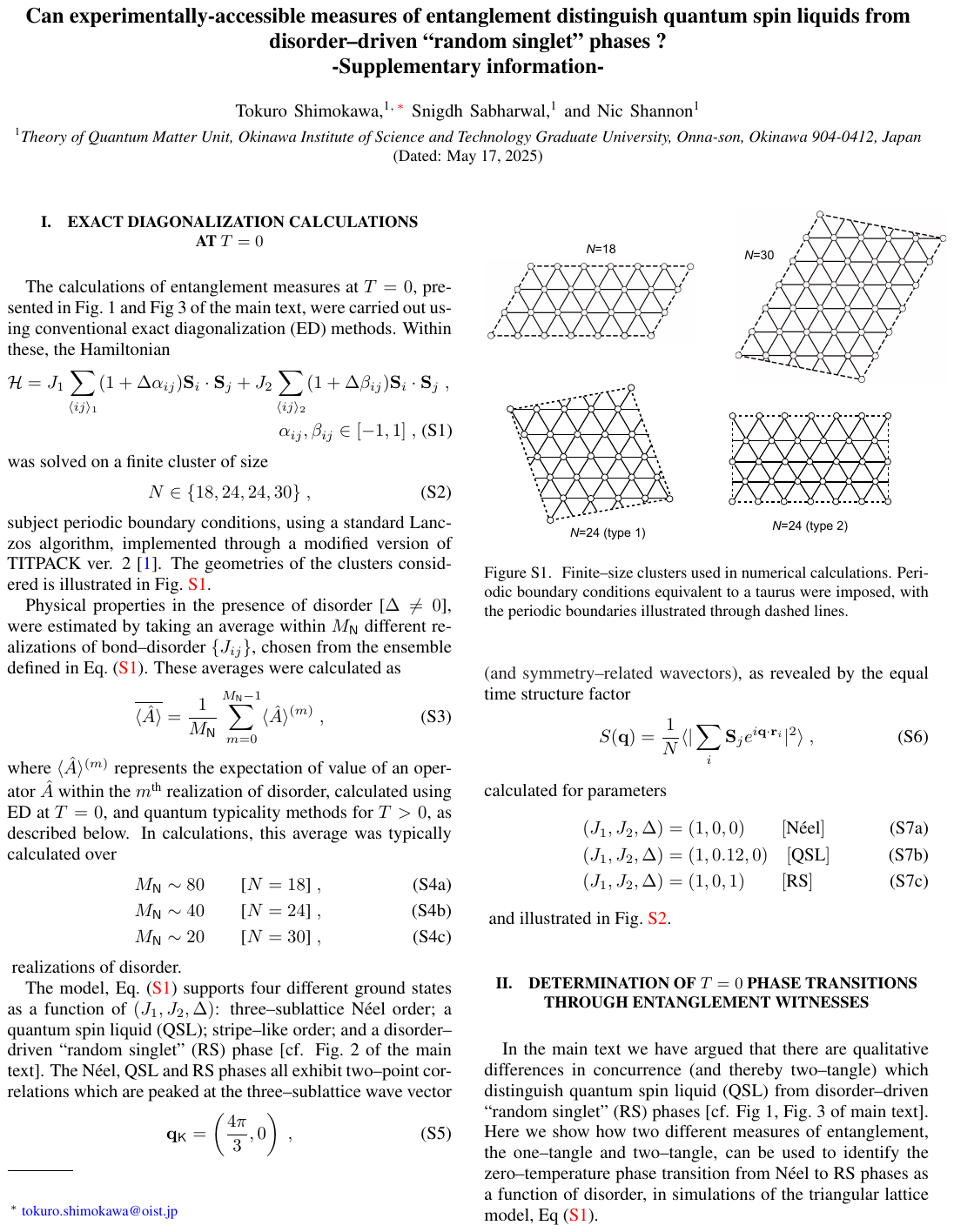}
    \includegraphics[page=2,scale=1.0, trim=18mm 0mm 0mm 15mm, clip]{supple.pdf}
    \includegraphics[page=3,scale=1.0, trim=18mm 0mm 0mm 15mm, clip]{supple.pdf}
    \includegraphics[page=4,scale=1.0, trim=18mm 0mm 0mm 15mm, clip]{supple.pdf}
    \includegraphics[page=5,scale=1.0, trim=18mm 0mm 0mm 15mm, clip]{supple.pdf}    
    \includegraphics[page=6,scale=1.0, trim=18mm 0mm 0mm 15mm, clip]{supple.pdf}
\end{center}

\twocolumngrid

 
 \end{document}